\documentclass[10pt]{IEEEtran}

\usepackage{booktabs}
\usepackage[utf8x]{inputenc}
\usepackage{graphicx}
\usepackage{subfigure}
\usepackage{placeins}
\usepackage{amssymb,amsmath,amsthm}
\usepackage{listings,color}
\usepackage[usenames,dvipsnames,svgnames,table]{xcolor}

\newcommand{\fhp}{FASTEST--3D}
\newcommand{\itac}{\textit{Intel Trace Analyzer and Collector}}
\newcommand{\lima}{LiMa}
\newcommand{\sumu}{SuperMUC}

\newcommand{\MBS}{\mbox{MByte/s}}
\newcommand{\GBS}{\mbox{GByte/s}}

\definecolor{light-gray}{gray}{0.95}

\title{Optimization of FASTEST-3D\\for Modern Multicore Systems}
\author{Christoph Scheit, Georg Hager, Jan Treibig, Stefan Becker, Gerhard Wellein}

\author{\IEEEauthorblockN{Christoph Scheit and Stefan Becker}\\
\IEEEauthorblockA{Institute of Process Machinery
and Systems Engineering,
University of Erlangen-Nuremberg\\
D-91058 Erlangen\\
Email: sh@ipat.uni-erlangen.de}\\[3mm]
\and
\IEEEauthorblockN{Georg Hager, Jan Treibig and Gerhard Wellein}\\
\IEEEauthorblockA{Erlangen Regional Computing Center (RRZE),
University of Erlangen-Nuremberg\\
D-91058 Erlangen\\
Email: georg.hager@fau.de}}

\begin{document}

\maketitle
\pagestyle{headings}
\begin{abstract}
  \fhp\ is an MPI-parallel finite-volume flow solver based on
  block-structured meshes that has been developed at the University of
  Erlangen-Nuremberg since the early 1990s.  It can be used to solve
  the laminar or turbulent incompressible Navier-Stokes equations. 
  Up to now its scalability was strongly limited 
  by a rather rigid communication infrastructure, which led
  to a dominance of MPI time already at small process counts.

  This paper describes several optimizations to increase the
  performance, scalability, and flexibility of \fhp. First, a
  node-level performance analysis is carried out in order to pinpoint
  the main bottlenecks and identify sweet spots for energy-efficient
  execution.  In addition, a single-precision version of the solver
  for the linear equation system arising from the discretization of
  the governing equations is devised, which significantly increases
  the single-core performance.  Then the communication mechanisms in
  \fhp\ are analyzed and a new communication strategy based on
  non-blocking calls is implemented.  Performance results
  with the revised version  show significantly
  increased single-node performance and considerably improved
  communication patterns along with much better parallel scalability.
  In this context we discuss the concept of ``acceptable parallel efficiency'' 
  and how it influences the real gain of the optimizations.
  Scaling measurements are carried out on a modern petascale
  system.  The obtained improvements are of major importance for the
  use of \fhp\ on current high-performance computer clusters and will
  help to perform simulations with much higher spatial and temporal
  resolution to tackle turbulent flow in technical applications.
\end{abstract}

\section{Introduction and related work}

The numerical solution of turbulent, unsteady flow problems requires
vast amounts of compute resources in order to get reasonable accuracy
and time to solution. This problem can be tackled in two complementing
ways: advanced numerical simulation algorithms and highly efficient, 
scalable implementations. 

This paper deals with \fhp, a finite volume solver based on
co-located, block-structured meshes. It originated from the Institute of Fluid Mechanics (LSTM) at the University
of Erlangen-Nuremberg is being developed since the early 1990s. Today different
versions of \fhp\ exist at the University of Erlangen-Nuremberg, the Technical University of
Darmstadt, and the University of Freiberg. The code is used to solve
the laminar or turbulent incompressible Navier-Stokes equations.  Time
evolution is based either on implicit schemes like Crank-Nicolson or
on explicit low-storage multi-stage Runge-Kutta time advancement
schemes. The linear equation system resulting from the discretization
of the governing equations is solved using Stone's~\cite{HLS68}
strongly implicit method (SIP), which is based on an incomplete $LU$
factorization. Several
$k$-$\epsilon$ and large-eddy simulation (LES) models are available 
for the simulation of turbulent flow.
Different coupling interfaces exists in order to simulate, e.g., fully
coupled fluid-structure interaction \cite{Glu02,FSSM+09},
one-way coupled acoustic interaction \cite{AEK+07}, and flow with
chemical reactions. Parallelization in \fhp\ is based
either on shared memory or on domain decomposition.
The latter approach is also used for the MPI domains in the hybrid
MPI/OpenMP version of \fhp. During the past years specific
parallelization and optimization strategies have been implemented in
order to improve the performance of \fhp\ on different high
performance systems including mixed/shared-memory parallelization
strategies~\cite{FD03,CBMB+02}. Most of those
optimizations have focused on vector computers with a low
number of processors and a relatively high workload per process,
hence the
communication overhead was relatively small. This is different on
massively parallel systems with a relatively small workload per process,
which leads to increased communication overhead, causing
poor parallel efficiency. In this paper we focus on the improvement of
single-core (and thus single-node) performance by work reduction
and the optional use of single-precision arithmetic, and on the
optimization of communication performance by using non-bocking
point-to-pint MPI functions, to the effect of a much-improved
massively parallel scalability.

This work is organized as follows: In Sect.~\ref{sec:testbed} we give
an overview of the hard- and software used and the benchmark cases.
Section~\ref{sec:code} shows an analysis of the \fhp\ code in terms
of function profiles and communication patterns. These results
are used in Sect.~\ref{sec:opt} for the optimization of 
single-core execution and the reduction of communication 
overhead. In Sect.~\ref{sec:perf} we demonstrate the achieved improvements
in performance and scalability on a petascale-class system.
Finally we give a summary and an outlook to future work in 
Sect.~\ref{sec:conc}.

\section{Test bed and benchmark cases}\label{sec:testbed}

\subsection{Test systems}

The scaling results in this paper were obtained on
\sumu~\cite{supermuc}, a federal system installed at Leibniz
Supercomputing Centre (LRZ) in Garching, Germany. \sumu\ is available
to scientists across Europe as a tier-0 system within PRACE
(Partnership for Advanced Computing in Europe). It consists of $18$
islands with $512$ compute nodes each.  A compute node comprises two
Intel Xeon E5-2680 (``Sandy Bridge'') eight-core processors, with an
achievable aggregate memory bandwidth of about 80\,\GBS\ (as obtained
with the standard STREAM benchmarks). The islands are fully
non-blocking QDR InfiniBand fat trees, with a 4:1 oversubscribed fat
tree across islands.

The sequential function profiles in Sect.~\ref{sec:profile} were taken on the
\lima\ cluster at Erlangen Regional Computing Center (RRZE) at the
University of Erlangen-Nuremberg. A compute node on \lima\ consists of two
Intel Xeon X5650 (``Westmere'') six-core processors, with an achievable 
aggregate memory bandwidth of about 40\,\GBS. 

The Intel compiler in version 12.1 and the Intel MPI library in
version 4.1 were used for all benchmarks on both systems.

\subsection{Test problems}
\begin{figure}[tb]
\centering
\includegraphics[width=0.95\linewidth]{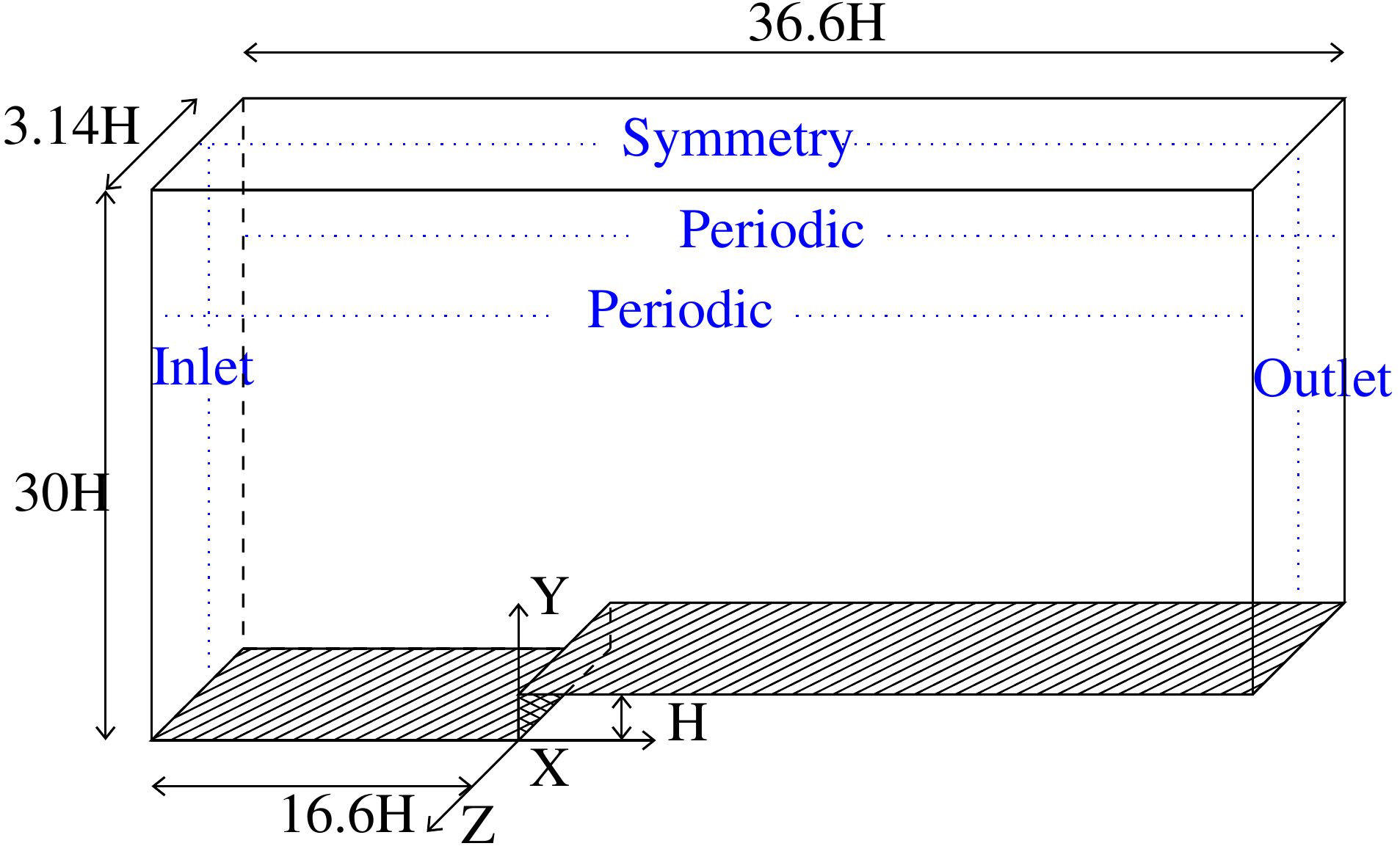}
\caption{Boundary conditions and domain for test case I, the forward--facing step}
\label{fig:domain_ffs}
\end{figure}
 %
%
\begin{figure}[tb]
\centering
\includegraphics[width=0.95\linewidth]{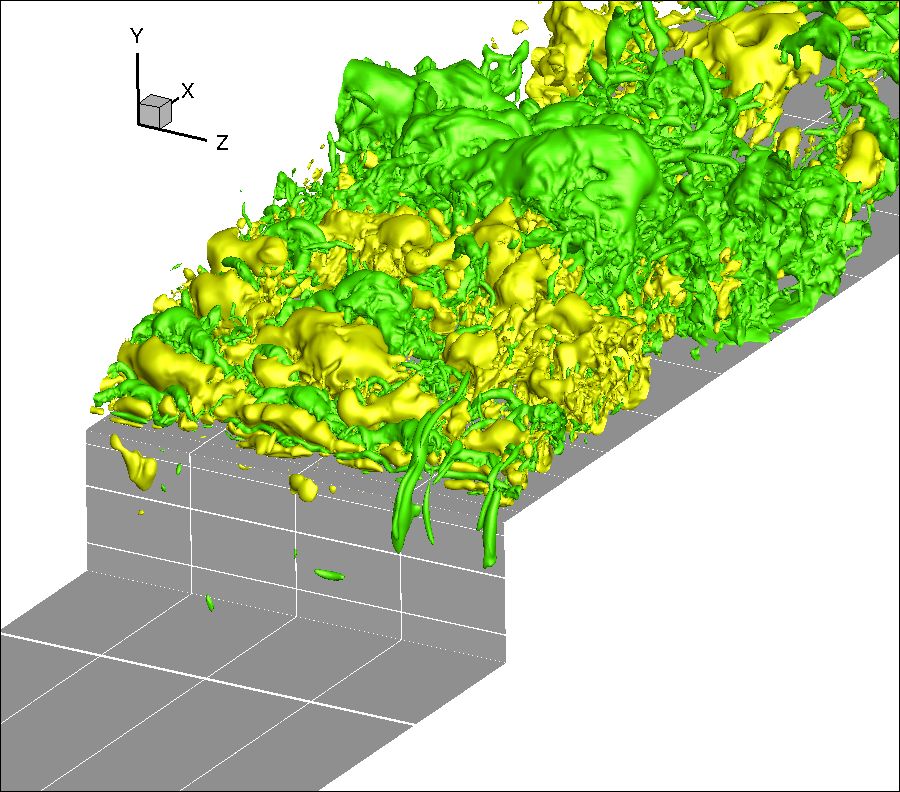}
\caption{Isosurfaces of turbulent pressure fluctuations for the flow over a forward-facing step computed via DNS; isovalues: 9/-9\;Pa}
\label{fig:isosurface_fluc_pres}
\end{figure}
Three different test problems were considered for the purpose of analysis:
\begin{itemize}
 \item[I] flow over a forward-facing step \cite{CSAE12}
 \item[II] turbulent flow in a plane channel, cf. Moser et al.~\cite{RDM99}
 \item[III] Taylor-Green Problem \cite{JKPM85}
\end{itemize}
For test case I, the simulation of the flow over a forward-facing step, we
used three
different grids consisting of a total number of $12.5\cdot 10^6$,  $280\cdot 10^6$,
and about $2200\cdot 10^6$ control volumes, respectively.
This case was used as an example of a real configuration
with grid sizes according to real-world large-eddy ($12.5$ Mio. control volumes) and
direct numerical simulations ($280$ and $2200$ Mio. control volumes).
Figure~\ref{fig:domain_ffs} shows the setup and Fig.~\ref{fig:isosurface_fluc_pres}
shows computed isosurfaces for this problem.
Test case II was used for validation of both the non-blocking communication strategy and
especially the single precision version of the linear equation solver. This test considers the fully
developed turbulent flow in a two-dimensional plane channel with periodic boundary
conditions in stream- and span-wise directions and no-slip wall
boundary conditions at the bottom and top wall. 
Test case III was also used for verification and to evaluate the single-socket
performance (see Sect.~\ref{sec:sopt} below), since this problem is perfectly
load-balanced including the boundary conditions (periodicity in $x$ and $y$ direction and
symmetry at bottom and top).

\section{Systematic code analysis of \fhp}\label{sec:code}
As a first step for the implementation of different communication,
parallelization and optimization strategies, a function profile of the
original version of \fhp\ was taken along with a basic investigation
of its requirements towards the hardware (such as memory bandwidth) and
its communication patterns.  For this analysis we used the
\textit{likwid} tools~\cite{JTGH+10}, the GNU profiler \textit{gprof},
and the \textit{Intel Trace Analyzer and Collector} \cite{entry-0e}.

\subsection{Function profile}\label{sec:profile}

Tables~\ref{tab:gnu_prof_imp} and
\ref{tab:gnu_prof_exp} show function profiles for the
serial version of \fhp\ using test case I (forward-facing step) with
implicit and explicit time discretization,
respectively. Only the ten subroutines consuming most of the total 
runtime are listed.
\begin{table}
\centering
 \begin{tabular}{llll}
 \toprule
time \% & self [s] & calls & name\\
 \toprule
  25.83 &   56.48  &   7500 & resforward3d\\
  22.53 &   49.26  &    945  & celuvw\\
  11.54 &   25.22  &  1200  & lu3d\\
   9.03  &   19.74  &    900  & celp2\\
   8.67  &   18.96  &   7500 & backward3d\\
   3.80  &     8.31  &    100  & calcp\\
   3.45  &     7.54  &   1916 & exvec\\
   2.96  &     6.48  &    300  & coefadd\\
   2.96  &     6.47  &    105  & caluvw\\
   2.32  &     5.07  &    205  & calcdp\\
   \bottomrule
   \hfill
  \end{tabular}
\caption{Sequential function profile, implicit time discretization, test case I}
\label{tab:gnu_prof_imp}
\end{table}
\begin{table}
\centering
  \begin{tabular}{llll}
  \toprule
  time \% & self [s] & calls & name\\
  \toprule
  32.93  &   26.93  &   3600 &  resforward3d\\
 16.68   &  13.64  &    900   & flx1\\
 11.26  &   9.21   &  3600     & backward3d\\
  9.34   &  7.64    &  135       & celuvw\_exp\\
  7.74   &  6.33     & 300       & lu3d\\
  6.49   &  5.31    &    5         & calcp\_exp\\
  5.09   &   4.16   &   215      & calcdp\_exp\\
  1.77  &   1.45   &  1529      & exsca\\
  1.47   &  1.20    &    5         & runge\_kutta\\
  1.22   &  1.00    & 1800      & flxnb1\\
  \bottomrule
  \hfill
 \end{tabular}
 \caption{Sequential function Profile, explicit time discretization, test case I}
 \label{tab:gnu_prof_exp}
\end{table}
In addition, Table~\ref{tab:gnu_mpi_prof} shows the profile 
for the explicit time discretization using six MPI processes on one
socket of \lima, to obtain a profile representative for
parallel execution on the first bottleneck (the socket) without
the adverse effects of communication overhead. 
The routines to solve the linear
equation systems arising from the discretization of the
Navier-Stokes equation are \textit{resforward3d}, \textit{lu3d} and
\textit{backward3d}. These routines together consume between
$45\,\%$ and $55\,\%$ of the total execution time. This fraction
depends on the settings for the linear equation solver and on the time
discretization scheme used: In the case of an explicit time
discretization only one equation for the pressure correction has to
be solved, while in the implicit time discretization
equation systems for the three Cartesian components of the momentum
equations have to be solved in addition. 
Another factor is 
the linear equation solver, i.e., the number of iterations per
correction step. However, based on this analysis one can identify
the linear equation solver as the most time-consuming part for both
parallel and serial execution. Hence, any serial code optimization
should first try to improve the performance of this part of
\fhp. One issue which can not be seen from the profile tables is that
the arrays used to store the coefficients of the linear
equation system are reused and overwritten by other, unrelated subroutines of
\fhp. Hence, even if the coefficients for the equations do not
change (as for the pressure correction equation in the
explicit time discretization), the coefficients have to be
recalculated. Since the explicit time discretization is of special
interest for DNS and LES, a redesign of the solver should avoid
unnecessary recalculation of the coefficients.  Finally, it should be
mentioned that the equation system for the pressure correction
equation is symmetric, but this symmetry has not been exploited so
far.
\begin{table}
\centering
         \begin{tabular}{llll}
        \toprule
 time \% &  self [s]   & calls &  name\\
 \toprule
 37.45   &    125.03 &     8100  &    resforward3d\_pp\\
 14.52   &     48.47  &     8100  &    backward3d\_pp\\
 11.26   &     37.59  &     1800  &    flx1\\
  9.01    &    30.07   &     645    &   calcdp\_exp\\
  6.78    &    22.65   &      15     &   calcp\_exp\\
  5.72    &    19.09   &     270    &    celuvw\_exp\\
  5.06    &    16.89   &     900    &    lu3d\_pp\\
  1.85    &     6.19    &   2400    &    setval\\
  1.60    &     5.33    &     15      &    runge\_kutta\\
  1.45    &     4.85    &   3600    &    flxnb1\\
  \bottomrule
  \hfill
  \end{tabular}
  \caption{Function profile using six MPI processes on one socket of \lima\ for
  explicit time discretization of test case I}
  \label{tab:gnu_mpi_prof}
\end{table}

Another possibility to gain more insight is the use of
likwid-perfctr~\cite{JTGH+10}, a simple command-line tool to measure
hardware performance metrics on x86 processors under Linux.
It also features a lightweight API which enables restricting the measurements
to certain parts of the code.
This allows, e.g., to get the memory bandwidth used while executing a specific
loop in the code. The result for a serial run of \fhp\ is shown in
Tab.~\ref{tab:likwid_membw}. In principle, by this analysis a piece of code
can be identified as being either memory bound or compute bound. Note that a single core
cannot saturate the memory bandwidth of a socket on modern multicore chips
even when running strongly memory-bound code~\cite{hager:cpe13}, for which
\fhp\ is a typical example. It can be seen from
Tab.~\ref{tab:likwid_membw} that the overall memory
bandwidth is in the range of $6$\,\GBS\ using the serial
version, with individual routines drawing between $2.5\,\GBS$ and $13\,\GBS$. 
The latter is also roughly the maximum bandwidth obtained by a simple
sequential streaming benchmark such as STREAM.
With $6$ Threads on one socket of \lima, an average bandwidth of over
$15$\,\GBS\ for the whole application can be reached, which is close
to the limit of $20$\,\GBS. Multiplying the values of the memory bandwidth in
Tab.~\ref{tab:likwid_membw} with the number of threads, most of the
measured code sections would easily exceed the maximal available
memory bandwidth. As a consequence, \fhp\ shows the typical saturation
behavior of a bandwidth-limited code on a multicore chip. See Sect.~\ref{sec:sopt}
below for some further analysis.
\begin{table}
\centering
    \begin{tabular}{lc}
    \toprule
                    Routine & Mem. BW [\MBS]\\
                    \toprule
                    Fastest-3D & 6212\\
                    calcp\_exp & 10894\\
                    calcp\_exp (2) & 6728\\
                    flx1 & 3845\\
                    calcdp\_exp & 12570\\
                    calcdp\_exp(2) & 13334\\
                    lu3d & 4204\\
                    backward3D & 7025\\
                    resforward3D & 7116\\
                    celuvw\_exp & 2490\\
                    \bottomrule
                    \hfill
                \end{tabular}
\caption{Per-routine analysis of memory bandwidth for the serial version (test case I) 
  on \lima}
\label{tab:likwid_membw}
\end{table}

\subsection{Communication pattern}
\begin{figure}
 \centering
 \subfigure[\label{fig:itac_bl_overview}Overview of total amount of communication time vs. computation time]{\includegraphics[width=0.95\linewidth]{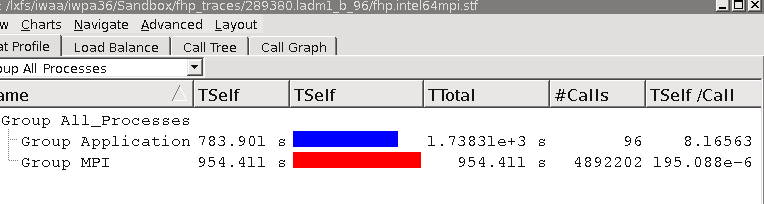}}\\
 \subfigure[\label{fig:itac_bl_timeline}Event time-line for a short time interval]{\includegraphics[width=0.95\linewidth]{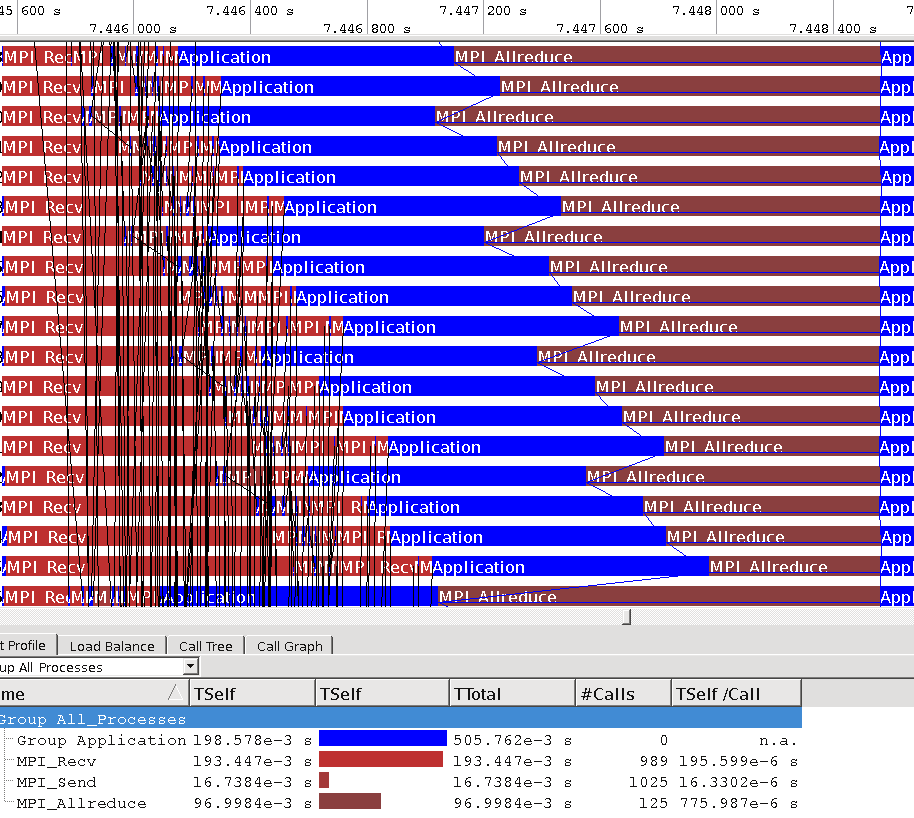}}\\
 \subfigure[\label{fig:itac_bl_ungroup_mpi}Distribution of communication time on different MPI functions]{\includegraphics[width=0.95\linewidth]{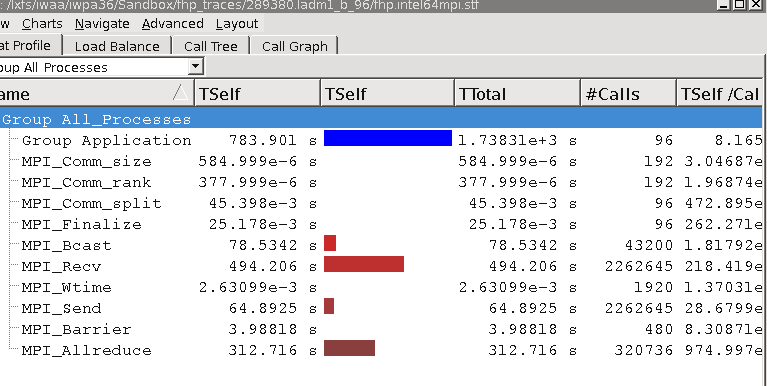}}
 \caption{ITAC analysis: Overview, event timeline, and time spent per MPI routine using one MPI process per node on the \lima\ cluster for test case II and 96 processes}
 \label{fig:itac_bl_screenshot}
\end{figure}

An analysis of the communication pattern has been performed using
\itac\ (ITAC). 
Typical screen shots of a run with the original code for test case II 
are shown in
Fig.~\ref{fig:itac_bl_screenshot} for a total number of 96 processes.
A single process per node was run in this experiment so that intra-socket
saturation effects could be ruled out.
As can be seen from Fig.~\ref{fig:itac_bl_overview} more time
is spent for communication (or rather within the MPI library)
than for computation. The number of control
volumes  (CVs) per process for this problem is $65536$, which is a
realistic number for a production run of \fhp, given enough parallel
resources. In a production environment one would require a minimum
acceptable parallel efficiency of $50\,\%$ (single node baseline) 
at problem sizes of approximately $50000$ CVs per process.
The parallel efficiency is defined as
\begin{equation}
 \varepsilon_{p}(N) = \frac{T_{s}}{NT_{p}(N)}~,
\label{eq:par_eff1}
\end{equation}
where $T_s$ and $T_p(N)$ are the execution times on a single node and on $N$ nodes,
respectively. Since the simulations can not always
be run on a single node due to the problem not fitting
into main memory, an alternative version for the parallel
efficiency will also been used below, where the baseline is
not a single node but $M$ nodes:
\begin{equation}
 \varepsilon_{p,M}(N) = \frac{T_{p}(M)}{NT_{p}(NM)}~,
\label{eq:par_eff2}
\end{equation}
Using this
definition, the base configuration already contains a certain amount
of (inter-node) parallel overhead. 

Based on the measurements obtained from ITAC, the parallel efficiency for
this configuration is already below the acceptable limit of $50\,\%$.
From Fig.~\ref{fig:itac_bl_timeline} it can be seen that the blocking
MPI\_Recv and MPI\_Send functions are used for point-to-point
communication between adjacent sub-domains/processes. While some
processes return relatively quickly from the call to MPI\_Recv, others
spend much longer in the call until a matching MPI\_Send has been issued. 
This leads to a partial serialization of
the communication, since the blocking calls to send/receive functions
are issued sequentially on each process, and
successive calls can not be issued until the current call returns.
This kind of partial communication serialization is a well-known 
pattern~\cite{GHGW10} that may be overcome by using 
non-blocking point-to-point send/receive calls
(MPI\_Isend/MPI\_Irecv). 
Also a large part of the communication time is lost in the
collective function MPI\_Allreduce. This is a consequence of the artificial
load imbalance caused by the serialization described above.
Moreover, reduction operations are mainly used in \fhp\ 
in order to compute the residual. By keeping those evaluations
to a minimum, the overhead from  MPI\_Allreduce can be reduced 
substantially.

\section{Code optimization}\label{sec:opt}
Using the results of the code analysis and the conclusions drawn from
it, a revised version of \fhp\ was developed.  In this section we give a
short overview of the major changes that were implemented to
improve the single-core performance and the scaling
properties of \fhp\ on multicore systems.
The
correctness of the improved code is shown for the important case
of a turbulent, fully developed channel flow.
\subsection{Improving the single-core performance}\label{sec:sopt}
From the measurements of the memory bandwidth and the runtime
profiles it was concluded that a relatively large part of the
overall runtime was spent solving the linear equation system and that
\fhp\ is a mostly memory-bound application. 
This is true for both explicit and implicit
time discretization, but the focus will be in the following on the
explicit time discretization, which is of special interest for this
work and direct numerical simulations performed with \fhp\ at the
moment. Three possibilities to improve the performance of the explicit
time discretization have been identified:
\begin{enumerate}
\item avoiding unnecessary re-computation of the coefficients of the
  pressure correction equation or the incomplete factorization matrices (ILU)
  constructed in Stone's implicit method~\cite{HLS68}
\item exploiting the symmetry of the pressure correction equation
\item using single precision to solve the linear equation system
\end{enumerate}
Regarding the first point, the use of an implicit time discretization scheme was
a very common use case for \fhp, and hence three equation systems resulting for
the momentum and one equation system resulting for the pressure correction have
to be solved. This is done in a segregated fashion in \fhp. By this
the coefficient arrays can be shared by all equations and thus are
overwritten by each other. For the explicit time discretization this
is not the case, since only one equation system is considered. Furthermore,
the coefficients of this equation system do not change and need to be calculated
only once. Consequently this is also true for the ILU factorization. Once computed,
the resulting decomposition can be used for all iterations and time steps without
the need to recompute it again. This fact was not exploited in the original
version of \fhp.

Concerning the second point, only the equation system for the pressure
correction equation is symmetric, since the convective operator in the
momentum equation is discretized using a deferred correction method and
a blend of central and upwind interpolation schemes yielding non-symmetric
equation systems. Hence, exploiting this property is most advantageous for
the explicit time discretization and the adoption for non-symmetric equation
systems has still to be investigated but is not the focus of the present work.
Taking advantage of this symmetry in the pressure correction equation is
expected to be beneficial, since the subroutine in question, \textit{resforward3d},
is memory bound.

The third measure implemented to improve the single core performance is
beneficial for both implicit and explicit time discretization without restrictions and
reduces the amount of data which has to be loaded while solving the
equation system. Since the rest of the algorithm is still performed
in double precision, an additional overhead for data conversion is
generated, slightly reducing the possible gain in performance.

\begin{figure}
\centering
\includegraphics*[width=0.9\linewidth]{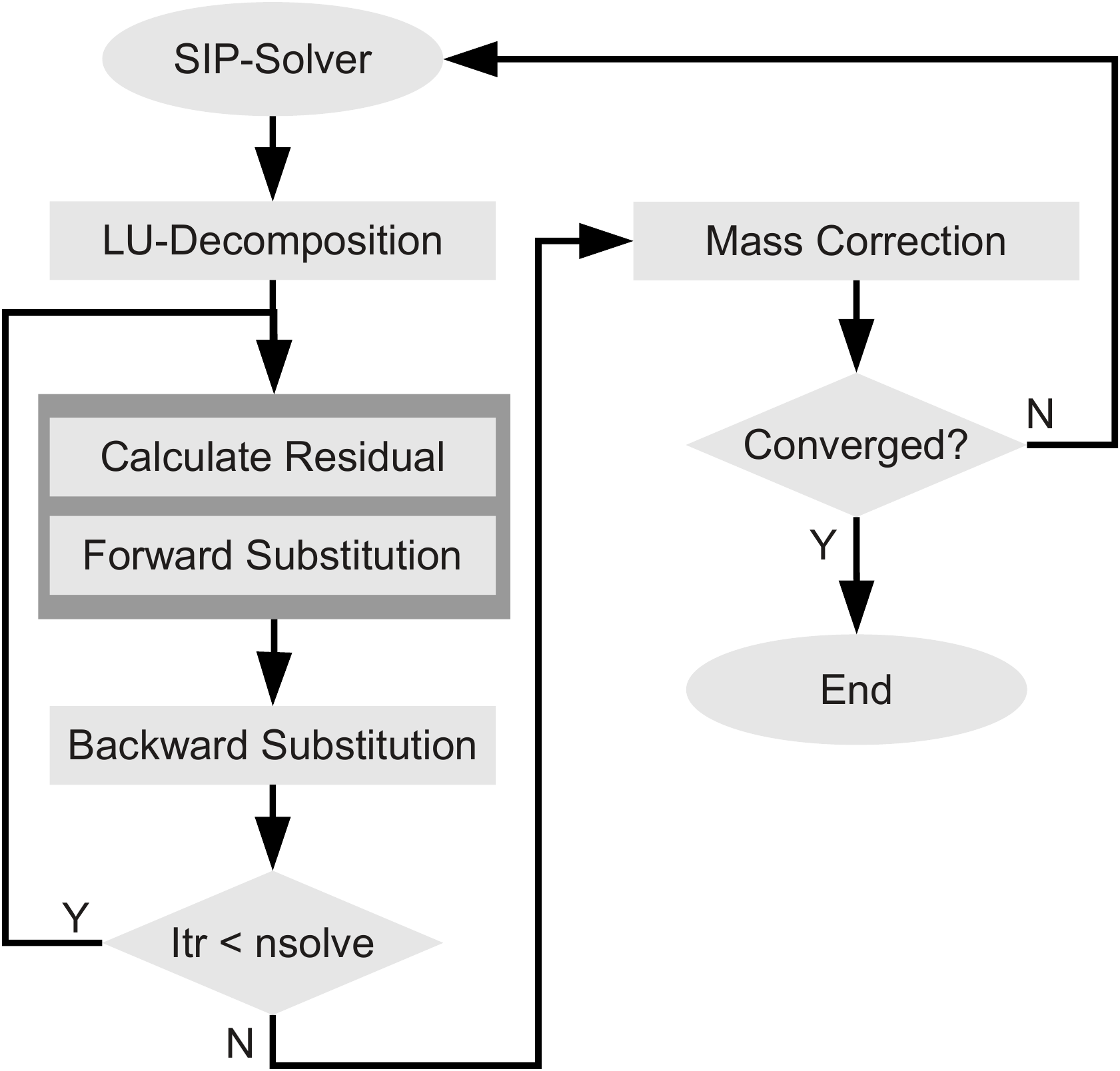}
 \caption{Original version of SIP solver}
 \label{fig:sip_solver}
\end{figure}

\begin{figure}
\centering
\includegraphics[width=0.9\linewidth]{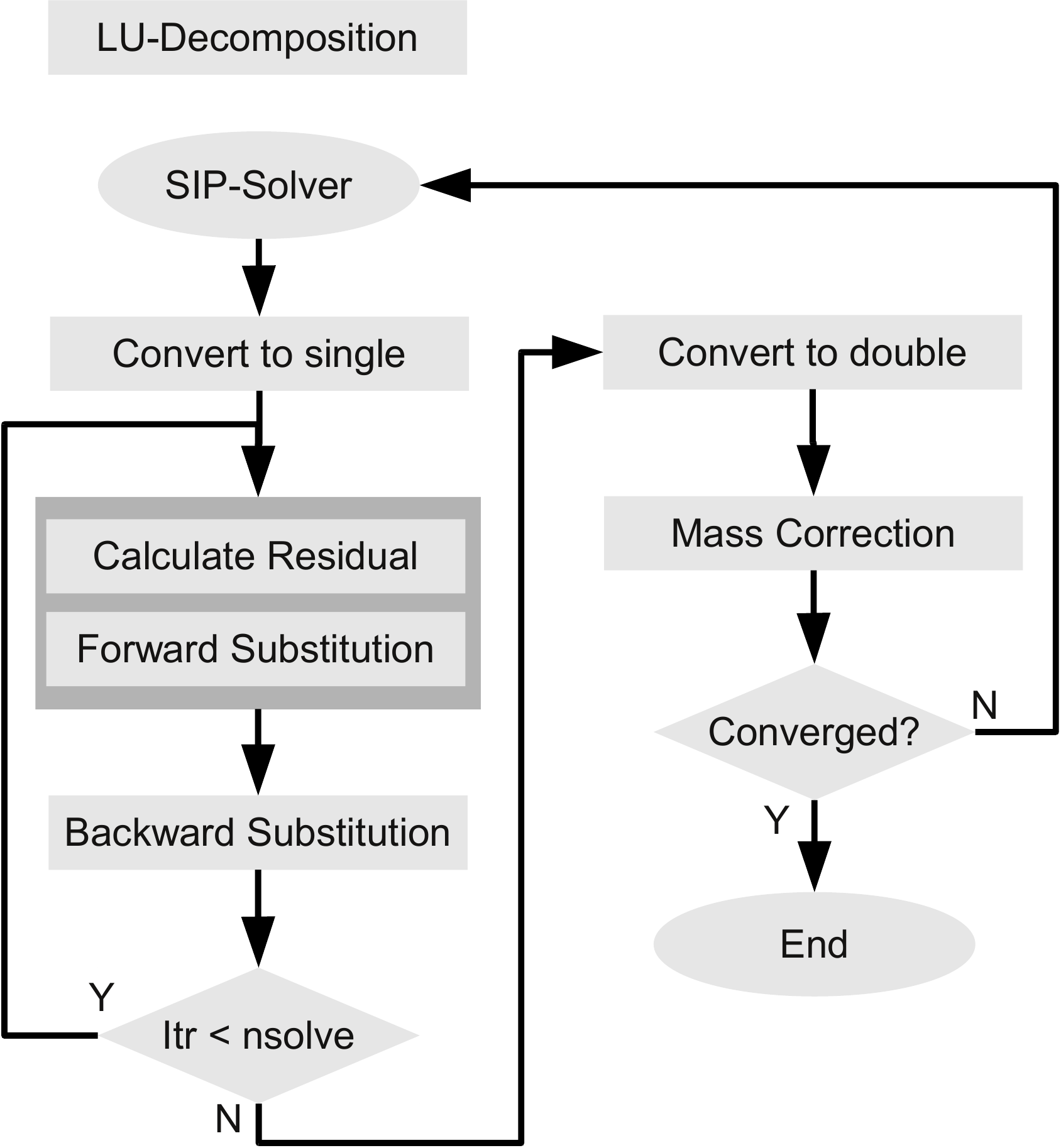}
 \caption{Optimized version of SIP solver}
 \label{fig:sip_solver_single}
\end{figure}

All of the above strategies were implemented in a redesign of the linear
equation solver and have changed the way the solver had to be interfaced.
Figures~\ref{fig:sip_solver} and~\ref{fig:sip_solver_single} show
the structure and call sequence of the SIP solver before and
after the redesign. In the original version, 
a call to the SIP solver always implied  a call to the subroutine for ILU
factorization. Then the linear equation system was relaxed until a given
number of iterations was reached. Finally, the pressure and velocity field
was corrected using the computed pressure correction and the outer
loop was checked for convergence, i.e., if the error in mass conservation
is below a given threshold.

The implementation of the new version was
based on a modular concept. The solver allocates and deallocates
memory for the coefficients, which are not overwritten any more and
thus do not need to be recomputed in between. Furthermore, the computation
of the ILU factorization is done in advance to the actual solver routine.
Before performing forward and backward substitution, the right hand side
and solution vectors have to be converted from double to single precision.
After the relaxation, only the solution vector is converted back to double
precision for the subsequent pressure and velocity correction.

In order to take the symmetry of the linear equation system into
account, the calculation of the residual, which is part of the forward
substitution step, was changed so that only the values of the east,
north and top coefficients are used (see Listing~\ref{list:res_calc}).
\lstset{language=Fortran, tabsize=2}
\begin{center}
\begin{tabular}{l}
  \begin{lstlisting}[caption=Residual calculation for a symmetric system matrix,label=list:res_calc, backgroundcolor=\color{light-gray}]
do k = 2, nk-1
  do i = 2, ni-1
    do j = 2, nj-1
      res(j,i,k) =
          ae(j,i,k) * fi(j,i+1,k)
       + ae(j,i-1,k) * fi(j,i-1,k)
       + an(j,i,k) * fi(j+1,i,k)
       + an(j-1,i,k) * fi(j-1,i,k)
       + at(j,i,k) * fi(j,i,k+1)
       + at(j,i,k-1) * fi(j,i,k-1)
       + su(j,i,k) - ap(j,i,k) * fi(j,i,k)
    enddo
  enddo
enddo
  \end{lstlisting}
\end{tabular}
\end{center}

\begin{figure}
 \includegraphics[width=0.95\linewidth]{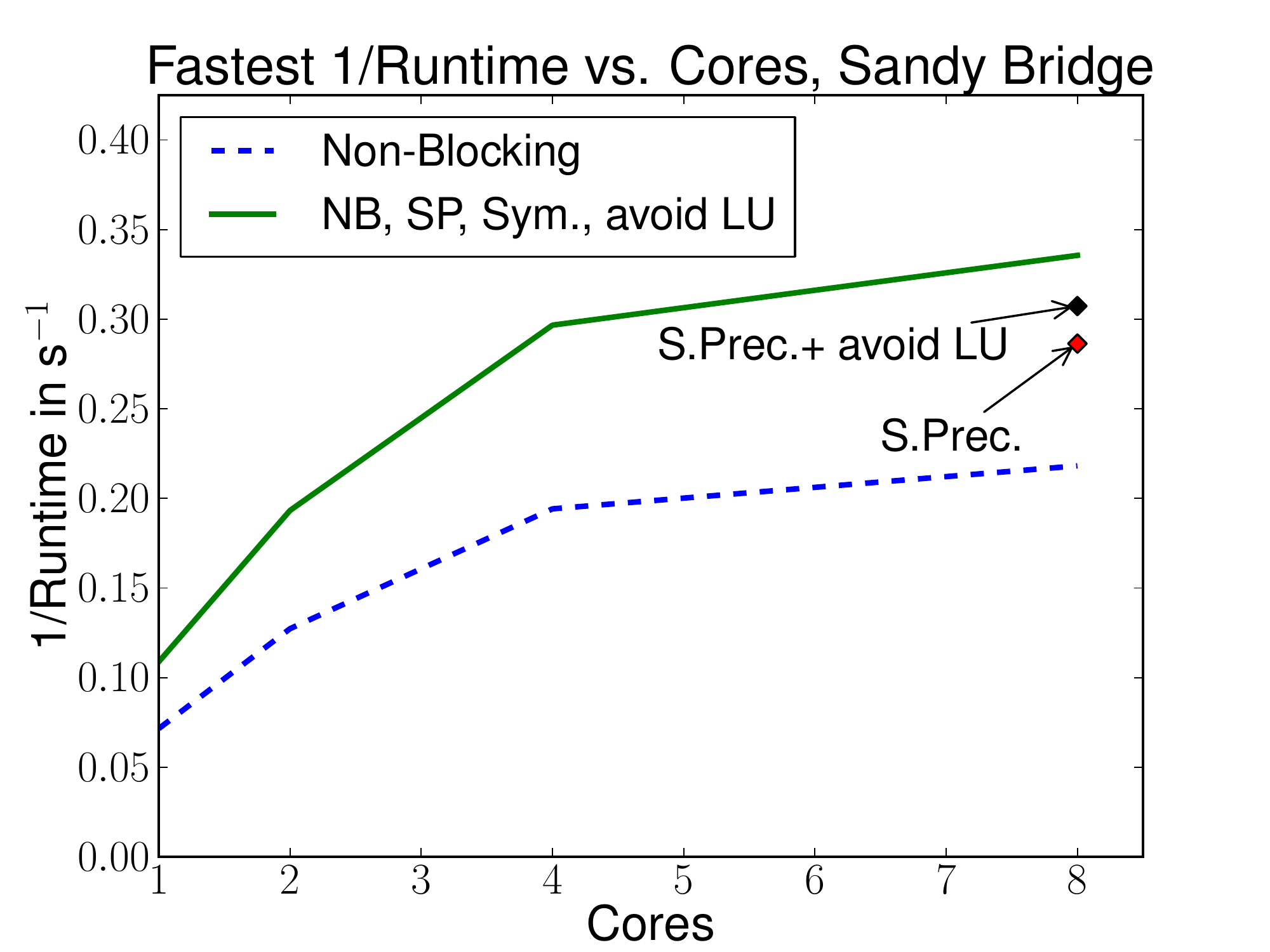}
 \caption{Scaling on one socket of an Intel Xeon Sandy Bridge node, non-blocking code
 version with additional code optimizatinos applied (test case III)}
 \label{fig:single_socket_perf_sandy_bridge}
\end{figure}

The effect of the different single-core optimizations has been measured on
one socket of \sumu. The dashed line in Fig.~\ref{fig:single_socket_perf_sandy_bridge}
shows the inverse runtime per timestep of the code version using non-blocking communication
(see the next section) 
and the original version of the SIP solver. The solid line shows the performance of the
non-blocking code version in combination with all single core optimizations applied in the
new version of the SIP solver. To evaluate the influence of the single core optimizations,
additional measurements for the same case running with eight MPI processes were taken. They
are shown with two additional scattered data points in Fig.~\ref{fig:single_socket_perf_sandy_bridge}.
By performing the relaxation of the equation system using single
precision arrays, a reduction in runtime by $25$\,\% is achieved. This is in good agreement with
the function profile shown in Tab.~\ref{tab:gnu_mpi_prof}, where the two subroutines
\textit{resforward3d} and \textit{backward3d} consume about $52$\,\% of the overall runtime
at double precision.
Reducing the size of the data set by a factor of two leads to a speedup of two
for those memory-bound routines. Avoiding the recomputation of the ILU factorization yields
another reduction by $7$\,\%, again in agreement with the function profile
(see Tab.~\ref{tab:gnu_mpi_prof}) measured on \lima. 
The last change, exploiting the symmetry in the
system matrix, improves runtime by another $5$\,\%. The solid line in 
Fig.~\ref{fig:single_socket_perf_sandy_bridge} combines all optimizations.

%

Only a slight increase
in performance can be gained using eight instead of four cores on one socket. However, this
also indicates the \fhp\ is not completely memory bound, since some parts of the code still
benefit from the additional number of cores.

\subsection{Improving communication performance}
The data exchange at block boundaries in \fhp\ is
based on transfer tables. These tables are set up in a preprocessing
step and stored in a binary file which is read by \fhp\ after a
simulation is started. Each process stores only information about the
send/receive operations it is involved in. Also no separate tables for
process-local and remote data transfer are present. Finally, all
communication routines are wrappers around the basic MPI routines,
providing a consistent interface no matter if the underlying
communication library is MPI, no communication (in case of a serial
run), or Parallel Virtual Machine (PVM). In addition, \fhp\ uses
different exchange routines for scalars, vectors, and tensors.
Figure~\ref{fig:flowchart_blocking_exchange} shows the basic
flow of the blocking data exchange in the original version:
\begin{figure}
 \includegraphics[width=0.95\linewidth]{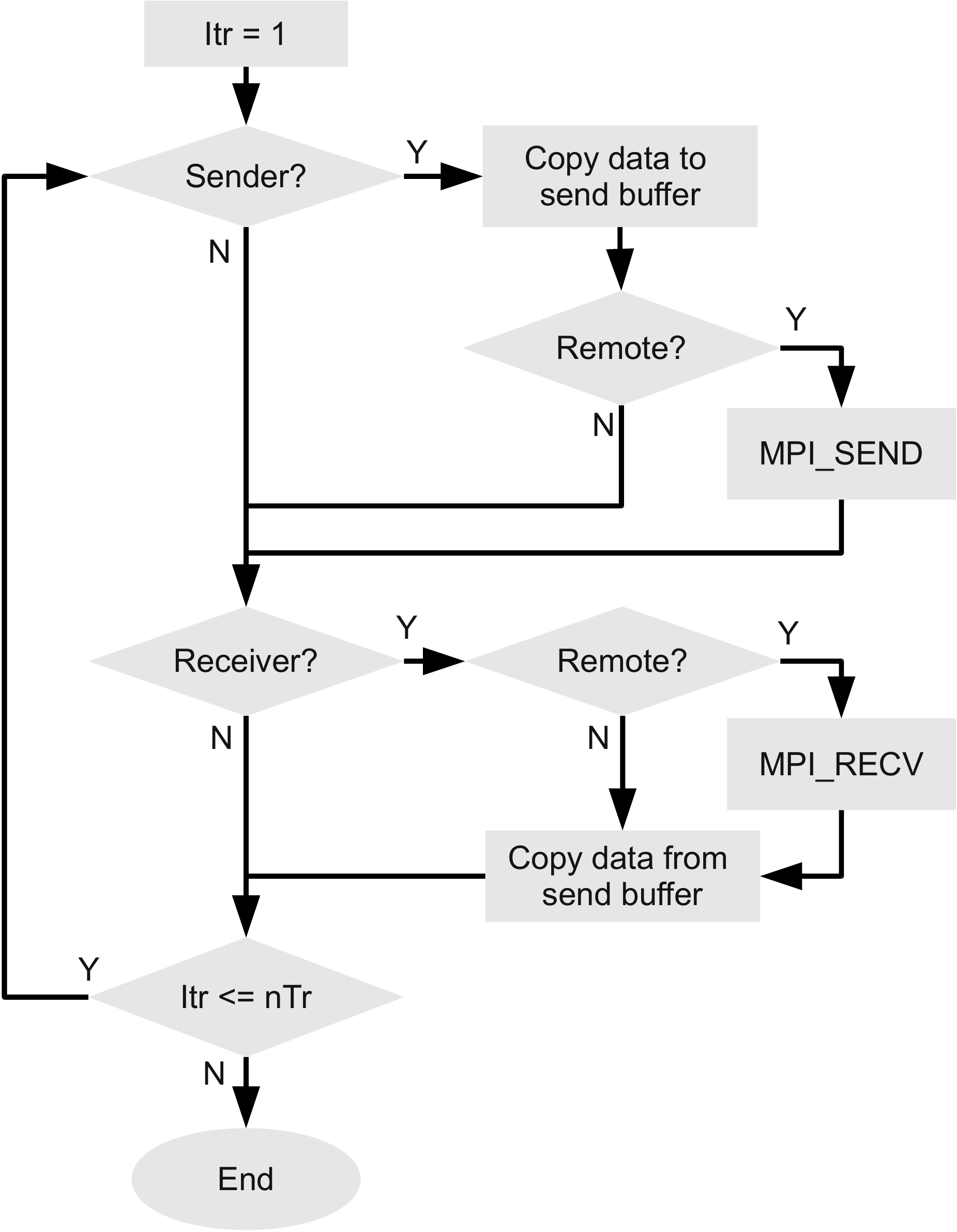}
 \caption{Blocking data exchange at block boundaries in \fhp.}
 \label{fig:flowchart_blocking_exchange}
\end{figure}
Each process performs a loop over all its patches
(block boundaries) that need to exchange data with their neighboring blocks,
regardless of whether the neighboring block is located on the same process. If the
patch between two neighboring blocks is on the same process, no MPI
routines are called and only a local copy operation is performed.
Otherwise the data is sent to the neighboring
process. Since MPI\_Send is a blocking call, the sending process has
to wait until the data is received at least for larger messages. 
This leads to a partial serialization, as described above,
and also applies to data copied to diagonal neighbors into
block edges and corner points.

\begin{figure}
 \includegraphics[width=0.95\linewidth]{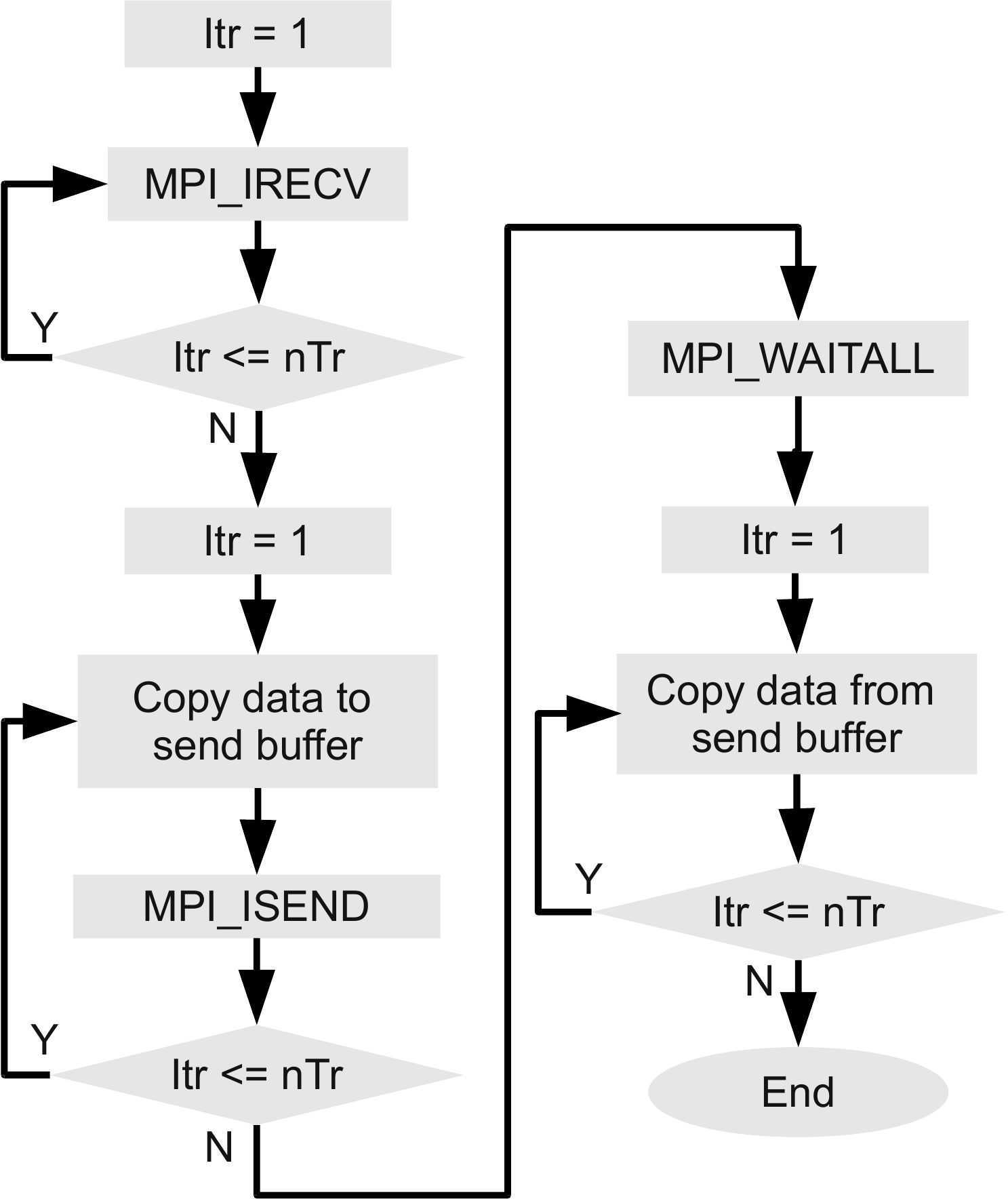}
 \caption{Non-Blocking data exchange at block boundaries in \fhp.}
 \label{fig:flowchart_non-blocking_exchange}
\end{figure}

In order to perform data exchange based on non-blocking MPI calls
a module was created that, after an initialization process,
sets up derived data transfer tables from the original 
tables by splitting these in data transfers between blocks on the
same process and remote data transfers for remote sends and
remote receives. For each kind of transfer an according exchange
routine was provided.  To ensure that all data is received by the
correct target process the MPI message tag is used, which is incremented
after each data transfer at block boundaries.  In addition we exploit the
property that if process \textit{A} sends two or more messages to process
\textit{B}, the messages arrive in the same order as they were sent.
As can be seen in Fig.~\ref{fig:flowchart_non-blocking_exchange}, first the calls to
MPI\_Irecv are made, than all send statements are issued, and finally 
MPI\_Waitall is called to ensure that all outstanding non-blocking
sends and receives are finished. In a final step
the data is copied from the receive buffers to the arrays which store
the variables. In contrast to the original algorithm no data which was
sent is available in the arrays which store the actual data since all
data remains in the receive buffer until all send and receive
statements have returned. Hence, the data at the block corner points
and block edges is always data from the previous exchange operation.
As long as all variables converge and a sufficient number of
operations is done, this is no problem since the difference between
variables from two consecutive exchange operations is very small
at convergence.  Also, the data at corners/edges is only needed for the
solution of the equation system for the diffusive fluxes on the right
hand side of the momentum equation. In order to have all data
correctly exchanged at the same time level, a three step approach
would be necessary, first exchanging corner points, then edges, and
finally the inner points of block patches. After each exchange step a
call to MPI\_Waitall would also be required, resulting in complex
exchange routines and preprocessing as well as in additional overhead.
%
\subsection{Verification}
\begin{figure}%
\centering
\includegraphics[width=0.95\columnwidth]{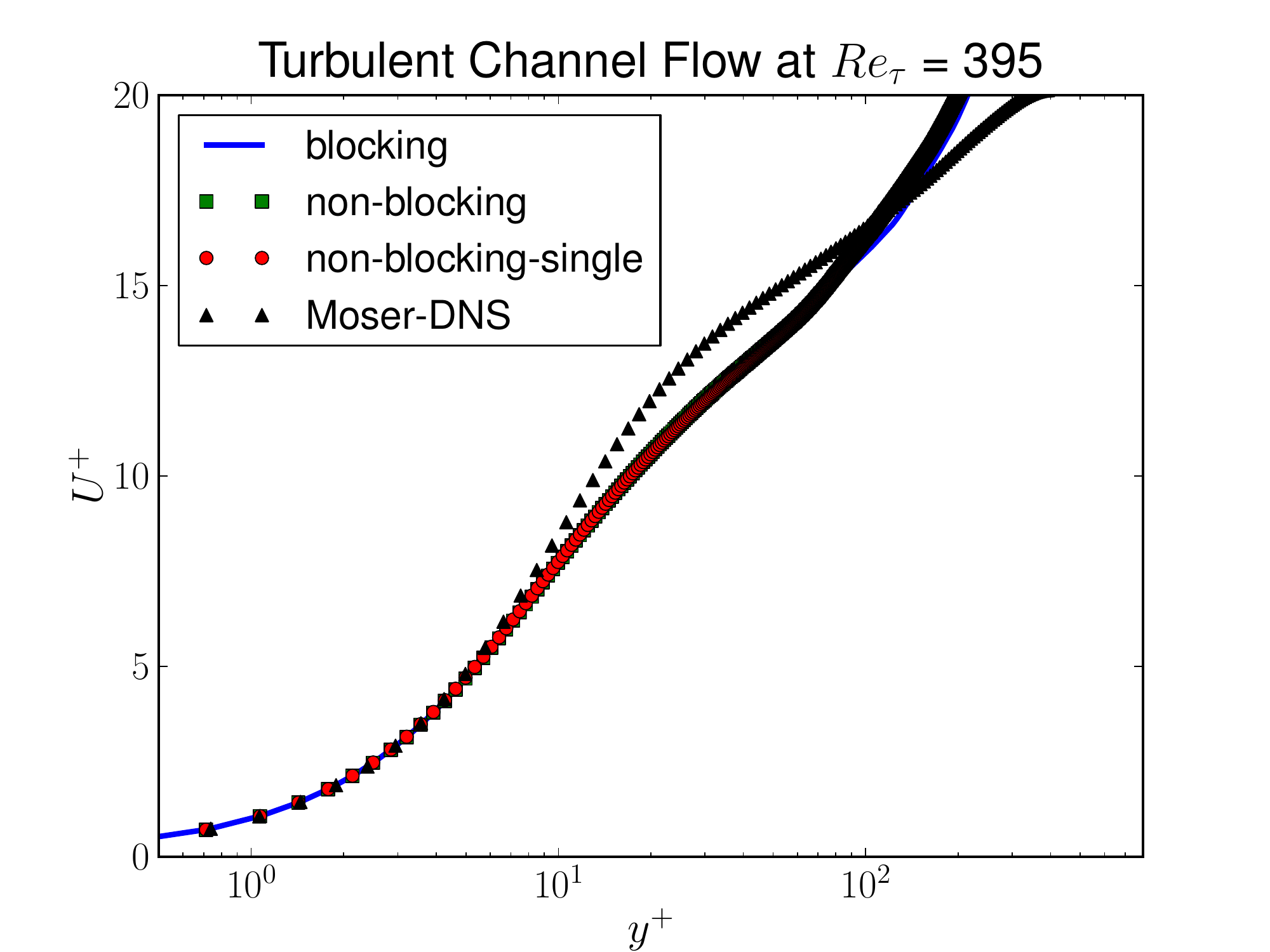}%
\caption{Comparison of original, non-blocking, and non-blocking version with
single core performance optimizations of \fhp\ with respect to DNS data of 
Moser et al.~\cite{RDM99} }%
\label{fig:comp_turb_channel_flow}%
\end{figure}

In order to verify the correctness of the new implementation, test
problem II is compared to the original implementation.
All three versions (original, with non-blocking calls, and non-blocking plus
single-core optimizations) were used to compute the mean velocity profile for
a turbulent channel flow (see Fig.~\ref{fig:comp_turb_channel_flow}). The velocity
in $x$ direction ($U$) and the distance from the wall ($y$) are normalized
by the friction velocity $u_{\tau}=\mu \frac{\partial U}{dx}$ with dynamic viscosity
$\mu$, and the kinematic
viscosity divided by the friction velocity, respectively, which is indicated by the
superscript $+$. All three variants are very close to
each other and no clear advantage can be seen for either of them in terms
of the quality of the solution.
Compared to the DNS data of Moser et al.~\cite{RDM99} a small deviation
in the log-law region for all the three variants can be observed; this may be
due to the lower approximation order for the derivatives in the governing equations
compared to Moser et al.~\cite{RDM99}, who used highly accurate
spectral methods while keeping the same grid sizes. On the other hand, the
\fhp\ version with non-blocking calls and a single-precision solver
performs best using the same hardware with respect to the time to solution.
\section{Performance and Scaling Results}\label{sec:perf}

In this section we present selected strong scaling measurements on \sumu\
for different code
versions of \fhp\ with 16 processes per node (ppn). 
Results are reported for the reciprocal of the averaged time
spent per time step ($1/T_r$) with runtime per time step $T_r$ and
parallel efficiency following the definition in
Eq.~\eqref{eq:par_eff2}. Performance is plotted vs. the number of compute
nodes, since this is the smallest allocatable unit on any modern
cluster system.
All results
presented here are based on test case I, the flow over the forward facing
step, which is a real-world problem and not only useful for
benchmarking purposes.
\subsection{Scaling Results on SuperMUC}
\begin{figure}
\centering
\includegraphics[width=0.95\columnwidth]{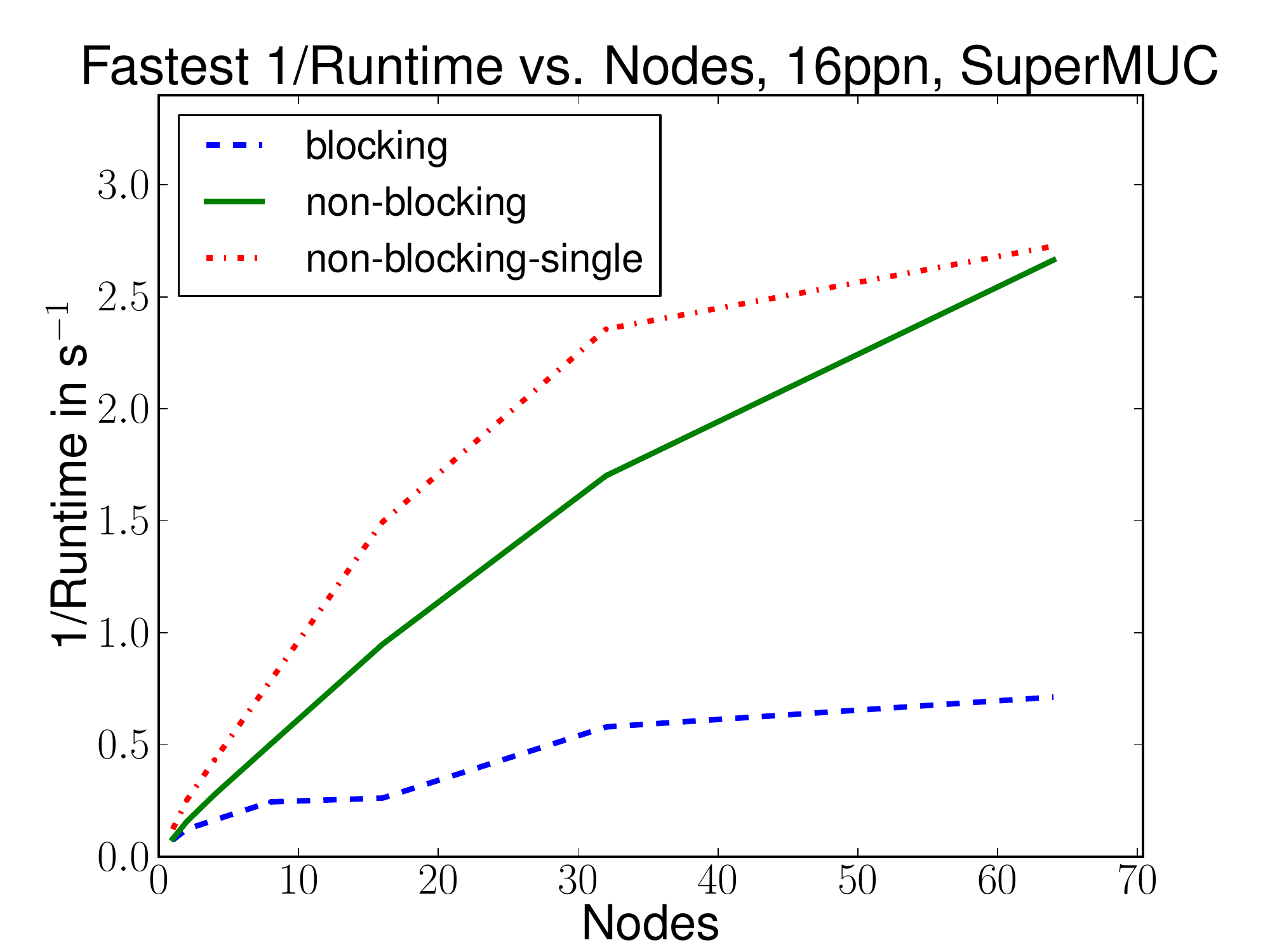}
\caption{Inverse runtime per timestep vs. nodes on SuperMUC, 16 ppn, test case I, $12.5\cdot10^6$\;CVs, strong scaling}
\label{fig:perf_supermuc}
\end{figure}

\begin{figure}
\centering
\includegraphics[width=0.95\columnwidth]{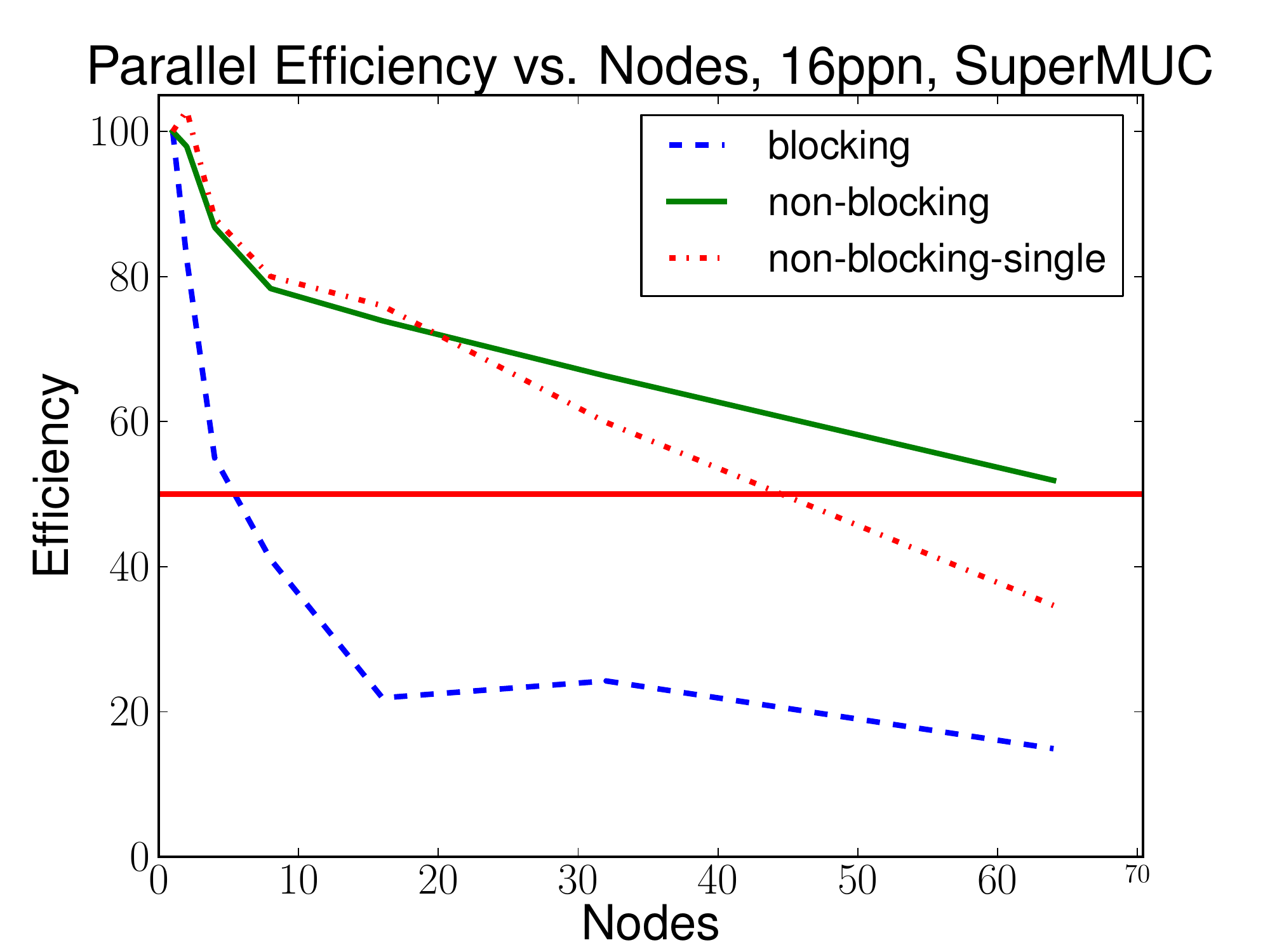}
%
\caption{Parallel efficiency vs. nodes on SuperMUC, 16 ppn, test case I, $12.5\cdot10^6$\;CVs, strong scaling}
\label{fig:eff_supermuc}
\end{figure}

Scaling measurements were performed
for domain sizes of $786432$ control volumes per process down to $12288$
for the coarsest grid, $387072$ down to $24192$ control volumes for the grid
consisting of $280$ Mio. control volumes and from $1.5$ Mio. down to $193536$
control volumes per process for the largest grid. The inverse of the averaged runtime
per timestep and the parallel efficiency are shown in Figures~\ref{fig:perf_supermuc} and
Fig.~\ref{fig:eff_supermuc} for the blocking and non-blocking versions of \fhp\ 
at the smallest grid size. Dashed lines correspond to the
original code version with blocking MPI calls and the original version of the SIP solver.
Solid lines denote the version with non-blocking MPI calls, and
dotted lines correspond to the code version using non-blocking communication and
the improved SIP solver.
%
As can be seen from Fig.~\ref{fig:perf_supermuc}, the best performance
is obtained for the non-blocking version with the improved SIP solver,
but
the non-blocking version with the original SIP solver shows better
scaling because the slower linear system solver spends more time
performing calculations between two subsequent communication steps.
Using the optimized version of the SIP solver along with non-blocking
communication, nearly the same performance as for the non-blocking
communication with the original SIP solver can be obtained using only
half of the resources ($32$ vs. $64$ nodes). Hence, if double precision
is not necessary, using the optimized SIP solver along with non-blocking
communication is the best choice in terms of overall cost since less resources
for nearly the same time to solution are needed. The horizontal solid line
in Fig.~\ref{fig:eff_supermuc} denotes the threshold of minimum acceptable
parallel efficiency ($50$\,\%). 

Obviously the optimized version of the SIP solver is not efficient 
beyond $32$ nodes at this small problem size.
It is remarkable that the original version of \fhp\ is not efficient at all even
for very small node counts ($\geq 8$), which is due to its severe communication
inefficiencies.
Comparing the performance at the limit of acceptable efficiency of
$50\,\%$, the new version performs about ten times better for the test
problem. We regard this as the more relevant gain, as opposed to the factor 
of $5$--$6$ achieved when comparing both versions at the same number
of nodes (but vastly different parallel efficiencies). Whether $50$\,\%
is the right threshold to apply is certainly debatable, but the overall
conclusions are largely independent of the actual limit.
%
%

\begin{figure}
\centering
\includegraphics[width=0.95\columnwidth]{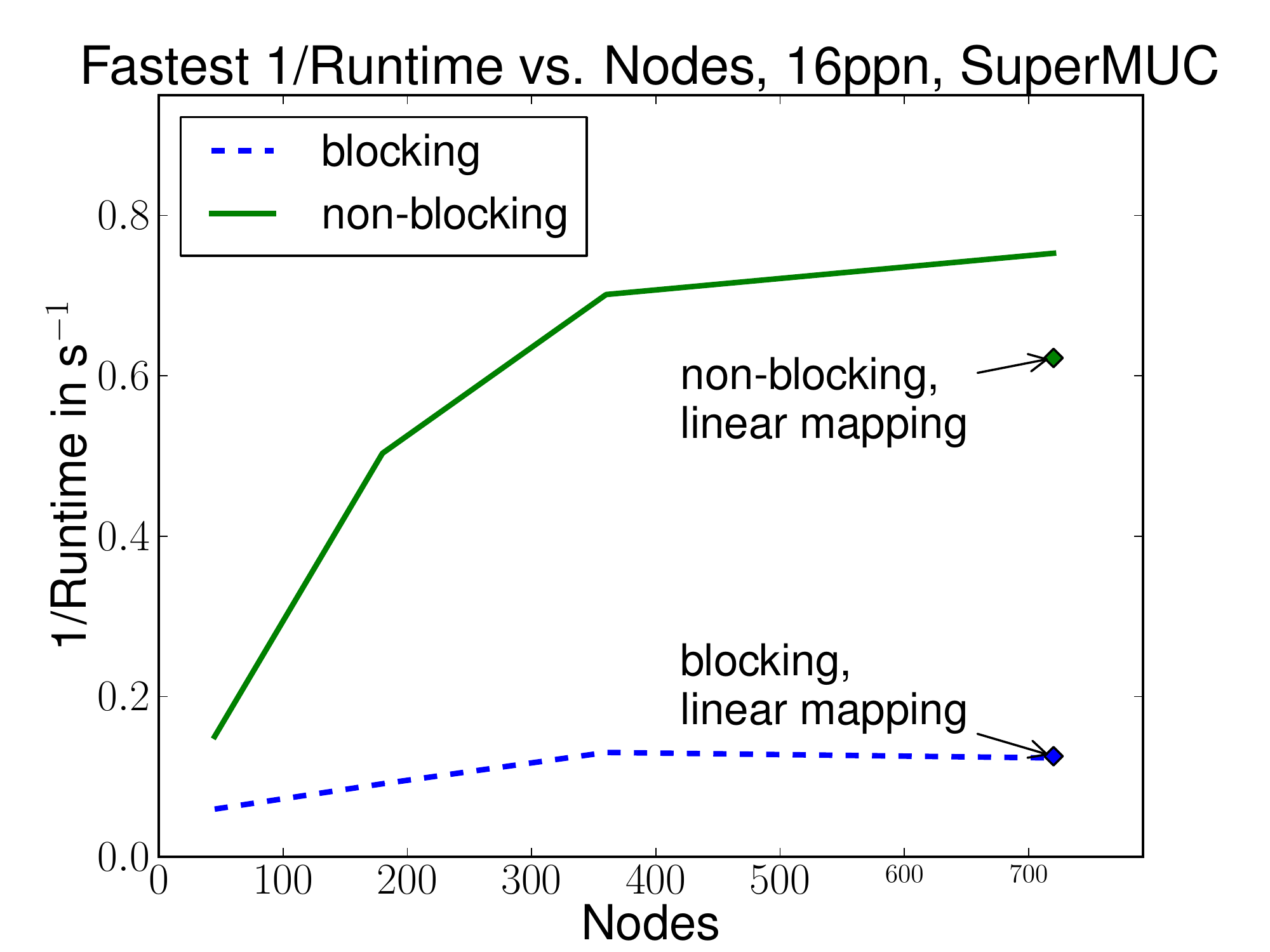}
\caption{Inverse runtime per time step vs. nodes on SuperMUC, 16 ppn, test case I, $280\cdot10^6$\;CVs, strong scaling}
\label{fig:perf_supermuc_full}
\end{figure}

\begin{figure}
\centering
\includegraphics[width=0.95\columnwidth]{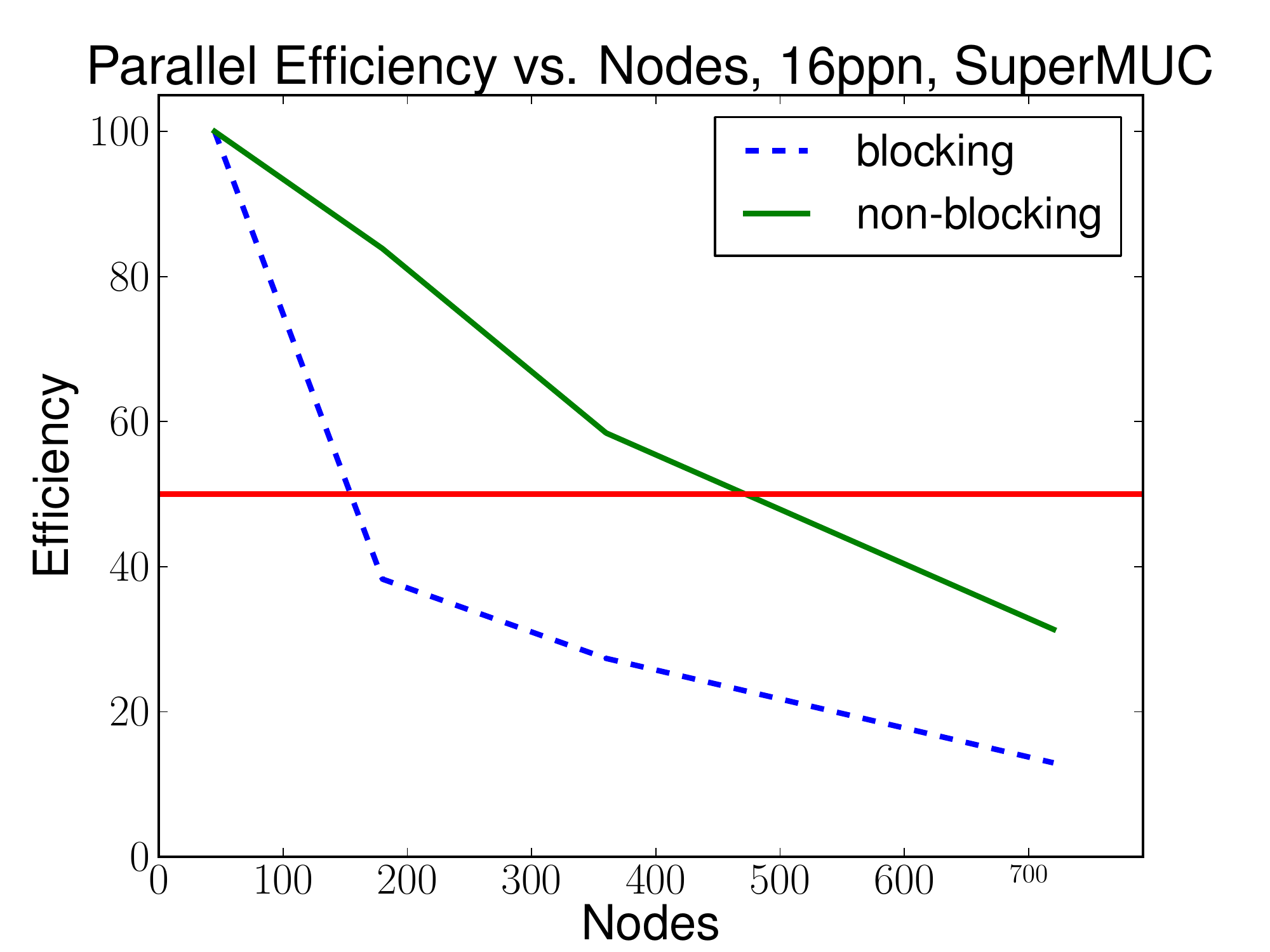}
\caption{Parallel efficiency on SuperMUC vs. nodes, 16 ppn, test case I, $280\cdot10^6$\;CVs, strong scaling}
\label{fig:eff_supermuc_full}
\end{figure}

In addition to the scaling runs at a moderate problem size,
additional measurements for an existing setup of a
direct numerical simulation (DNS) with approximately $280\cdot10^6$
control volumes were done on \sumu\, comparing the new version with
the original version of \fhp\ for a real simulation. Scaling results are shown
in Figures~\ref{fig:perf_supermuc_full} and~\ref{fig:eff_supermuc_full}.
Note that, in contrast to the measurements for the coarse grid,
scaling has been performed here for more than $512$ nodes, which means
that for the last data point in Figures~\ref{fig:perf_supermuc_full}
and~\ref{fig:eff_supermuc_full} two network islands have been used,
leading to a considerable drop in bisection bandwidth per node.
Results are reported for the code version using non-blocking communication
with the original SIP solver (solid line) and the original code version based on blocking
communication (dashed line). Again, considering the inverse runtime, the code
version using non-blocking communication shows much better parallel scaling.
While for the non-blocking version the parallel efficiency drops below $50$\,\%
for more than $500$ nodes, this is the case for the version using blocking calls already
for about $180$ nodes (see Fig~\ref{fig:eff_supermuc_full}). Comparing both
versions at approximately $50$\,\% parallel efficiency, the non-blocking version
is about $8$ times faster using only twice the amount of resources at the same level of parallel
efficiency. 

Fig.~\ref{fig:perf_supermuc_full} also shows two additional scattered data points
denoting the performance using a naive core-rank mapping which does not take the 
neighborhood relation between parallel
subdomains for MPI task placement into account. In case of the non-blocking
communication, neglecting the neighborhood relations causes a drop in performance
of about $15$\,\%. For the version using blocking communication this effect does not
play any role since the parallel efficiency for the last data point is already very low.
\begin{figure}
\centering
\includegraphics[width=0.95\columnwidth]{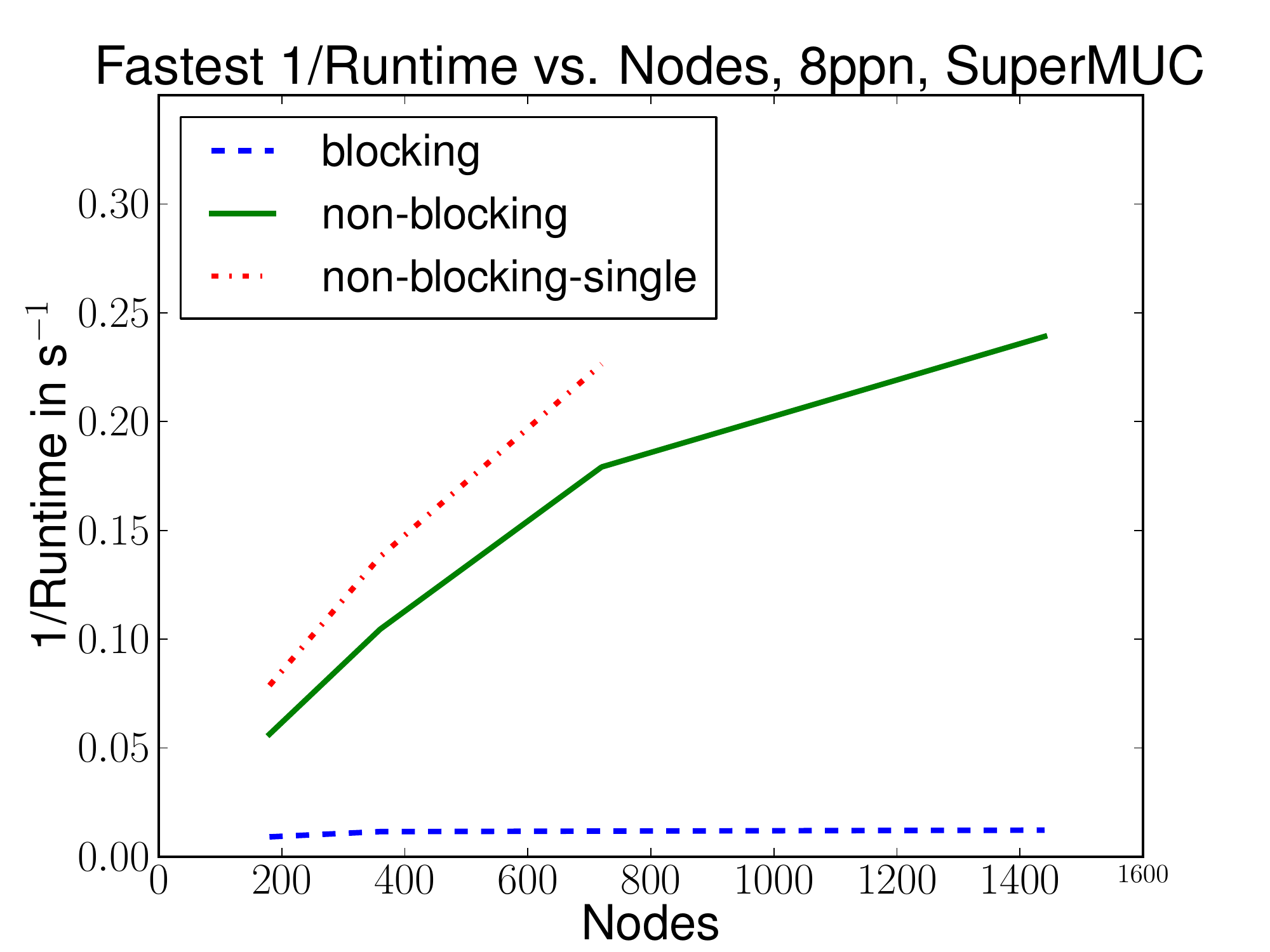}
\caption{Inverse runtime per time step vs. nodes on SuperMUC, 16 ppn, test case I, $2.2\cdot10^{9}$\;CVs, strong scaling}
\label{fig:perf_supermuc_2G}
\end{figure}
\begin{figure}
\centering
\includegraphics[width=0.95\columnwidth]{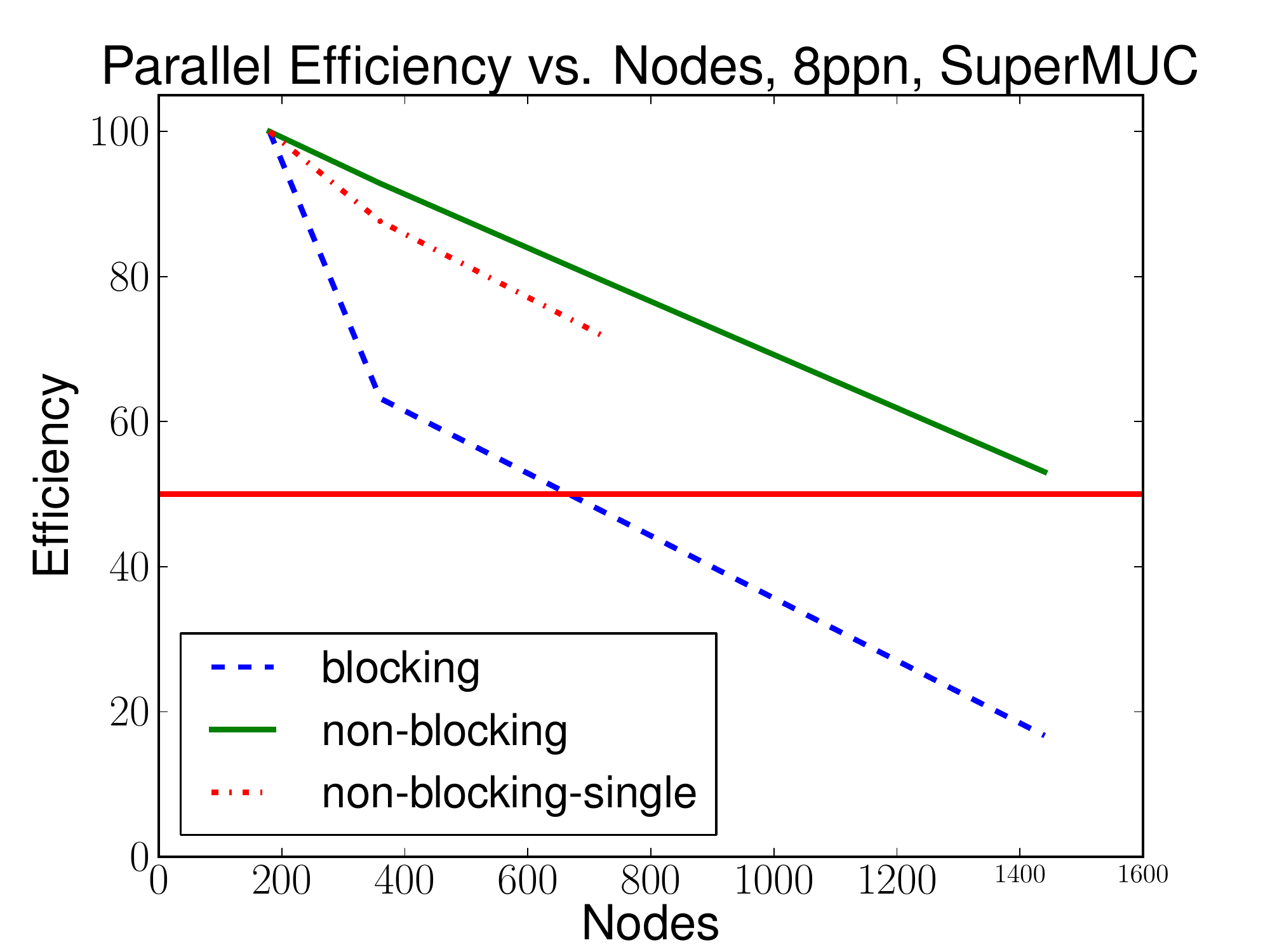}
\caption{Parallel efficiency on SuperMUC vs. nodes, 16 ppn, test case I, $2.2\cdot10^{9}$\;CVs, strong scaling}
\label{fig:eff_supermuc_2G}
\end{figure}

Finally, Figures~\ref{fig:perf_supermuc_2G} and \ref{fig:eff_supermuc_2G} show
performance and efficiency data for the largest set of $2.2\cdot 10^9$ CVs.
The optimized code version at double precision can scale efficiently
to node counts which are unreachable by the original code (beyond $1400$ nodes,
i.e., $22400$ cores). 
The single-precision SIP solver can still be of use even at these large node counts. Note that
the parallel efficiency shown in Fig.~\ref{fig:eff_supermuc_2G} is relative to a
$180$-node baseline due to the size of the problem.

\section{Conclusions and Outlook}\label{sec:conc}
We have implemented several optimizations in \fhp, a parallel
finite-volume flow solver based on co-located, block-structured
meshes, to improve its sequential performance and parallel scalability
on modern multicore systems. By work-avoiding strategies and employing
a single-precision linear solver, the single-node performance could be
improved by about $40$\,\%. At the MPI-parallel level, employing
non-blocking MPI for block boundary exchange resulted in a massive
improvement of strong parallel scalability, not because of overlapping
communication with computation but due to eliminating partial
communication serialization. For the purpose of comparing the cost of
computations we have introduced the concept of ``minimum acceptable
parallel efficiency.'' At an efficiency limit of $50$\,\% we could
achieve between $8$x and $10$x speedup over the original version,
depending on the problem size. The code in its optimized form is now
ready for massively parallel, strongly scaled simulations on current
high-performance cluster platforms.

Future work may include a more thorough study of the effects of
rank-core mapping on the scaling properties, and a detailed
analysis of the hybrid (MPI+OpenMP) version of \fhp, which 
is expected to show different communication and convergence
behavior compared to the pure MPI version studied here.

\section*{Acknowledgments}
This work was supported by the Bavarian Competence Network for
Scientific High Performance Computing in Bavaria (KONWIHR).

\FloatBarrier

\bibliographystyle{IEEEtran.bst}
\bibliography{literature_db}

\begin{thebibliography}{10}
\providecommand{\url}[1]{#1}
\csname url@samestyle\endcsname
\providecommand{\newblock}{\relax}
\providecommand{\bibinfo}[2]{#2}
\providecommand{\BIBentrySTDinterwordspacing}{\spaceskip=0pt\relax}
\providecommand{\BIBentryALTinterwordstretchfactor}{4}
\providecommand{\BIBentryALTinterwordspacing}{\spaceskip=\fontdimen2\font plus
\BIBentryALTinterwordstretchfactor\fontdimen3\font minus
  \fontdimen4\font\relax}
\providecommand{\BIBforeignlanguage}[2]{{%
\expandafter\ifx\csname l@#1\endcsname\relax
\typeout{** WARNING: IEEEtran.bst: No hyphenation pattern has been}%
\typeout{** loaded for the language `#1'. Using the pattern for}%
\typeout{** the default language instead.}%
\else
\language=\csname l@#1\endcsname
\fi
#2}}
\providecommand{\BIBdecl}{\relax}
\BIBdecl

\bibitem{HLS68}
{Herbert L. Stone}, ``Iterative {S}olution of {I}mplicit {A}pproximations of
  {M}ultidimensional {P}artial {D}ifferential {E}quations,'' \emph{SIAM J.
  Numer. Anal.}, vol.~5, no.~3, pp. 530--558, Sep. 1968.

\bibitem{Glu02}
M.~Gl\"uck, ``Ein {B}eitrag zur numerischen {S}imulation von
  {F}luid-{S}truktur-{I}nteraktion -- {G}rundlagenuntersuchungen und
  {A}nwendung auf {M}embrantragwerke,'' Ph.D. dissertation, Univeristy of
  Erlangen-Nuremberg, Germany, 2002.

\bibitem{FSSM+09}
{F. Sch\"afer}, {S. M\"uller}, {T. Uffinger}, {S. Becker}, {J. Grabinger}, and
  {M. Kaltenbacher}, ``Numerical and {E}xperimental {I}nvestigations on the
  {F}luid-{S}tructure-{A}coustic {I}nteraction of the {F}low {P}ast a {T}hin
  {F}lexible {S}tructure,'' \emph{AIAA Journal}, vol.~48, no.~4, pp. 738--748,
  2009.

\bibitem{AEK+07}
I.~Ali, M.~Escobar, M.~Kaltenbacher, and S.~Becker, ``Time {D}omain
  {C}omputation of {F}low {I}nduced {S}ound,'' \emph{Comput. Fluids}, vol.~37,
  no.~4, pp. 349--359, May 2008.

\bibitem{FD03}
{Frank Deserno}, ``Basic {O}ptimisation {S}trategies for {CFD}-{C}odes,''
  Regionales Rechenzentrum Erlangen, Tech. Rep., 11 Jul. 2003.

\bibitem{CBMB+02}
{C. Bartels}, {M. Breuer}, {K. Wechsler}, and {F. Durst}, ``Computational
  {F}luid {D}ynamics {A}pplications on {P}arallel-{V}ector {C}omputers:
  {C}omputations of {S}tirred {V}essel {F}lows,'' \emph{Computers \& Fluids},
  vol.~31, no.~1, pp. 69--97, Jan. 2002.

\bibitem{supermuc}
\BIBentryALTinterwordspacing
``{SuperMUC} petascale system.'' [Online]. Available:
  \url{http://www.lrz.de/services/compute/supermuc}
\BIBentrySTDinterwordspacing

\bibitem{CSAE12}
{C. Scheit}, {A. Esmaeili}, and {S. Becker}, ``Direct {N}umerical {S}imulation
  of {F}low over a {F}orward-{F}acing {S}tep - {F}low {S}tructure and
  {A}eroacoustic {S}ource {R}egions,'' in \emph{Conf. {M}od. {F}luid {F}low},
  {J. Vad}, Ed., vol.~2, Sep. 2012, pp. 891--898.

\bibitem{RDM99}
{R. D. Moser}, {J. Kim}, and {N. N. Mansour}, ``Direct numerical simulation of
  turbulent channel flow up to {R}e\_tau = 590,'' \emph{Phys. Fluids}, vol.~11,
  no.~4, pp. 943--945, 1999.

\bibitem{JKPM85}
{J. Kim} and {P. Moin}, ``Application of a {F}ractional-{S}tep {M}ethod to
  {I}ncompressible {N}avier-{S}tokes {E}quations,'' \emph{J. Comp. Physics},
  vol.~59, pp. 308--323, 1985.

\bibitem{JTGH+10}
{Jan Treibig}, {Georg Hager}, {Gerhard Wellein}, and {Michael Meier},
  ``L{IKWID} {P}erformance {T}ools,'' \emph{Innovatives Supercomputing in
  Deutschland}, vol.~1, no.~8, pp. 50--53, 2010.

\bibitem{entry-0e}
``{h}ttp://software.intel.com/en-us/intel-trace-analyzer.''

\bibitem{hager:cpe13}
\BIBentryALTinterwordspacing
G.~Hager, J.~Treibig, J.~Habich, and G.~Wellein, ``Exploring performance and
  power properties of modern multicore chips via simple machine models,''
  submitted. [Online]. Available: \url{http://arxiv.org/abs/1208.2908}
\BIBentrySTDinterwordspacing

\bibitem{GHGW10}
{Georg Hager} and {Gerhard Wellein}, \emph{Introduction to {H}igh {P}erformance
  {C}omputing for {S}cientists and {E}ngineers}.\hskip 1em plus 0.5em minus
  0.4em\relax Chapman \& Hall/CRC Press, 2 Jul. 2010.

\end{thebibliography}

\end{document}